\documentclass[showpacs,preprint,preprintnumbers,amsmath,amssymb,showkeys]{revtex4}

\usepackage{graphicx}% Include figure files
\usepackage{dcolumn}% Align table columns on decimal point
\usepackage{bm}% bold math

\begin{document}

\preprint{gr-qc/XXX}

\title{Reconstruction of source location in a
 network of gravitational wave interferometric detectors}

\author{Fabien Cavalier}
\email{cavalier@lal.in2p3.fr}
\author{Matteo Barsuglia}
\author{Marie-Anne Bizouard}
\author{Violette Brisson}
\author{Andr\'e-Claude Clapson}
\author{Michel Davier}
\author{Patrice Hello}
\author{Stephane Kreckelbergh}
\author{Nicolas Leroy}
\author{Monica Varvella}

\affiliation{Laboratoire de l'Acc\'el\'erateur Lin\'eaire, IN2P3-CNRS et
Universit\'e Paris-Sud 11,
Centre Scientifique d'Orsay, B.P. 34, 91898 Orsay Cedex (France)}

\date{July 14, 2005}

\begin{abstract}

This paper deals with the reconstruction of the direction of a gravitational wave source using the detection made by a network of interferometric
detectors, mainly the LIGO and Virgo detectors. 
We suppose that an event has been seen in coincidence using a filter applied on the 
three detector data streams.
Using the arrival time (and its associated error) of the gravitational signal in each detector, 
the direction of the source in the sky is computed using a $\chi^2$ minimization technique.
For reasonably large signals (SNR$>$4.5 in all detectors), the mean angular error between
the real location and the reconstructed one is about $1^{\circ}$. We also investigate the effect of the network geometry 
assuming the same angular response for all interferometric detectors. It appears that the reconstruction 
quality is not uniform over the sky and is degraded when the source approaches the plane defined by the three detectors.
Adding at least one other detector to the LIGO-Virgo network reduces the blind regions and in the case of 6 detectors,
a precision less than $1^{\circ}$ on the source direction can be reached for 99\% of the sky. 

\end{abstract}



\maketitle

\section{Introduction}

The LIGO and Virgo gravitational wave interferometric detectors are approaching their design
sensitivity \cite{LIGO},\cite{Virgo} and in the near future, coincidences between the three detectors
(LIGO-Hanford, LIGO-Livingston and Virgo) will be possible. In order to reconstruct the direction of the 
astrophysical sources in the sky, it is well known \cite{Tourrenc}
that a minimum of three detectors is mandatory even if 
an ambiguity remains between two positions symmetric with respect to the plane defined by the 3 detectors. 
The source direction is also provided by the coherent searches for bursts 
\cite{Gursel}, \cite{LAL1}, \cite{Sylvestre1} 
or coalescing binaries \cite{Jaranowski}, \cite{Bala}, \cite{Pai} where one of the outputs of the detection algorithm
is an estimation of the  source direction.

In this paper, we propose a method
for estimating the source position using only the arrival time of the gravitational signal in each detector.
The event detection is supposed to have been previously done by dedicated algorithms ([10-23] for bursts, [24-29] for coalescing binaries) 
and is not within the scope of this article.

The direction reconstruction is based on a $\chi^2$ minimization as described in Section II. This technique
can be easily extended to any set of detectors. Moreover, the method can be applied to
several types of sources (burst, coalescence of binary objects ...) as soon as an arrival time can be defined for the
event.

Section II also deals with the simulation
procedure which will be used in the following sections to evaluate the reconstruction quality in several 
configurations. Sections III and IV describe the 
performances of the LIGO-Virgo network first neglecting (III), then including (IV) the angular response of the detectors. 
In section V, we consider the addition of other gravitational wave detectors (supposing a similar sensitivity) and investigate their impact on the 
reconstruction.
In real conditions, systematic errors on arrival time are likely to exist and their impact on the reconstruction is tackled in Section 
VI.

\section{Modeling the reconstruction of the source direction}

\subsection{Arrival time of GW signals in interferometers}

Within a network of {\it n} interferometers, we suppose that 
each detector $D_i$ measures the arrival time $t_i$ of the gravitational wave.
Of course, the definition of the arrival time
depends on the source type and is a matter of convention, for example: peak value in the case of a supernova signal,
end of the coalescence for binary events. In the following, it is assumed
that all interferometers use consistent conventions.

The error on the arrival time, $\sigma_i$, depends on the estimator used and on the strength of the signal
in the detector $D_i$, strength (for a given distance and a given signal type) which is related to the antenna pattern 
functions (see \cite{LAL1} and references therein)
at time $t_i$. At that time, the antenna pattern depends on the longitude and the latitude 
of the detector 
location, as well as its orientation, the angle between the interferometer
arms, the sky coordinates $\alpha$ (Right Ascension) and $\delta$ (Declination) of the source, and the
wave polarization angle $\psi$. 

The timing uncertainty $\sigma_i$ can be parametrized by \cite{LAL2}:

\begin{equation}
  \label{eq:sigma_t}
  \sigma_i = \frac{\sigma_0}{(SNR_i)^{\zeta}}
\end{equation}

where $SNR_i$ is the measured SNR in detector $D_i$, $\sigma_0$ and $\zeta$ are constants depending on the detection
algorithm and the signal shape.
For example, a burst search with a 1-ms Gaussian correlator leads to $\sigma_0 \simeq 1.4$ ms 
and $\zeta=1$. Typically, for an SNR equal to 10, the error on the arrival time is a few tenth of milliseconds and 
weakly depends on $\zeta$ for SNR values between 4 and 10.

\subsection{$\chi^2$ Definition}

The $\it n$ measured arrival times $t_i$ and their associated errors $\sigma_i$ are
the input for the reconstruction of the source direction in the sky, 
direction defined by $\alpha$ and $\delta$.

In the 3-detector configuration, the angles ($\theta$ and $\phi$ ) of the source 
in the detector coordinate system (see Ref.\cite{Gursel} and \cite{Jaranowski} for exact definitions)
are given by \cite{Jaranowski}:

\begin{eqnarray}
\sin \theta &=& \frac{1}{a b_2} \sqrt{\Delta}\\ \nonumber
\cos \theta &=& \pm \frac{1}{a b_2} \sqrt{ (ab_2)^2 - \Delta}\\ \nonumber
\sin \phi &=& - \frac{b_2 (t_1 - t_2)}{\sqrt{\Delta}}\\ \nonumber
\cos \phi &=& \frac{a (t_1 - t_3) - b_1 (t_1 - t_2)}{\sqrt{\Delta}} \\
\mbox{with} \\ \nonumber
\Delta &=& ( b_1 (t_1 - t_2) - a (t_1 - t_3))^2 + (b_2 (t_1 - t _2 ))^2
\label{eq:jara_1}
\end{eqnarray}

where D$_1$ is placed at (0,0,0), D$_2$ at 
($a$,0,0) and D$_3$ at ($b_1$,$b_2$,0).

When performing a coherent analysis of the GW detector streams 
the position of the source in the sky is part of the output parameters, corresponding to the stream combination 
which maximizes the SNR. However, for a burst search, it is known that thousands of possible positions have to be tested
to obtain the solution \cite{LAL1}, \cite{Sylvestre1} or a least-square function involving the integration
of detector streams has to be minimized \cite{Gursel}. This minimization also implies the test of
hundreds of initial conditions in order to reach the right minimum.

Concerning coalescing binaries, it implies the definition of a five-parameter bank of filters
including the chirp mass, the three Euler angles and the inclination angle \cite{Jaranowski} of the orbital plane
or a three-parameter bank of thousands filters for the two source angles and the chirp time
\cite{Pai}.

In all coherent techniques, the extraction of the source direction is an heavy process imbedded in
the detection procedure.

In this paper, we propose a simpler approach where $\alpha$ and $\delta$ are found through a least-square minimization using 
separately triggered events 
obtained by
a coincidence search. We suppose that the detection is already performed applying suitable algorithms (matched filter for coalescing binaries,
robust filters for bursts).
The $\chi^2$ is defined by:

\begin{equation}
  \label{eq:chi2}
  \chi^2 = \sum_{i=1}^{n} \frac{\left(t_i - (t_0 + \Delta_i^{Earth}(\alpha,\delta) ) \right)^2}{\sigma_i^2}
\end{equation}
where $t_0$ is the arrival time of the gravitational wave at the center of the Earth and
$\Delta_i^{Earth}(\alpha,\delta)$ is the delay between the center of the Earth and
the i$^{th}$ detector which only depends on $\alpha$ and $\delta$.

The first advantage of this definition is that it deals with absolute times recorded by each detector rather than time differences
where one detector has to be singled out. Otherwise, the best choice for the reference detector is not 
obvious: the detector with the lower error on the arrival time, the detector which gives the larger time delays or the detector leading the best 
relative errors on timing differences ?
This definition leads to uncorrelated errors on fitted measurements.

The second advantage is that the network can be extended to any number of detectors and the addition of other detectors is 
straightforward.
Obviously, the method requires that the event is seen by all detectors.

The least-square minimization provides the estimation of $t_0, \alpha, \delta$ and the 
covariance matrix of the fitted parameters. When the number of detectors is greater than 3,
the $\chi^2$ value at the minimum can also be used as a discriminating variable, as the system is overconstrained.

\subsection{Simulation procedure for a 3-detector network}

A list of coincident events are defined by the three arrival times and their
associated errors. No detection procedure is performed in these simulations as stated before.

The simulation proceeds in two steps in order to study two coupled effects: antenna-patterns and location
with respect to the 3-detector plane.
The first step is a simplified approach: the antenna-pattern functions are ignored and the same error $\sigma_i$ 
is assumed for arrival times. The second step is more realistic: we assume the same sensitivity for each detector
and the signal strength is adjusted to have the mean (over the three detectors) SNR equal to 10. However, this implies that 
sometimes, due to the antenna-pattern, the signal is seen in a given detector with an SNR lower than 4.5, which remains
an acceptable threshold for a 
real detection. The same threshold equal to 4.5 will be used later to see the effect of a reasonable 
detection scheme on the angular reconstruction error. 
We evaluate the errors $\sigma_i$ for each arrival time $t_i$ using Equation \ref{eq:sigma_t} with 
$\zeta=1$. 

For a given simulation, the coordinates ($\alpha^{true}, \delta^{true}$) of the source are chosen
and the true arrival times $t_i^{true}$ on each detector are computed taken $t_0^{true}$ equal to 0 (it is obvious
that the timing origin can always been chosen such as $t_0^{true}=0$)

The measured arrival times $t_i^{measured}$ are drawn according to a Gaussian distribution 
centered on $t^{true}_i$ and of width $\sigma_i$.
The simulated values $t_i^{measured}$ are then used as inputs for the least-square minimization.
As the minimization is an iterative procedure, some initial values for the 
parameters have to be given. $t_0$ is initialized by the average of
$t_i^{measured}$. For the angles,
it appears that the initial values for the angles have no influence on the 
minimization convergence and a random direction is adequate.
In the case of three interferometers, it is well known that there is a twofold ambiguity for the direction in the sky
which can lead to the same arrival times in the detectors. These two solutions are symmetric with respect to 
the 3-detector plane. In order to resolve the ambiguity, a fourth 
detector is needed. For the evaluation of the reconstruction accuracy, only the solution closest to the source is retained.

\section{Source direction reconstruction: 3-detector network geometry}
\label{source}

\begin{figure}
  \includegraphics[width=10cm]{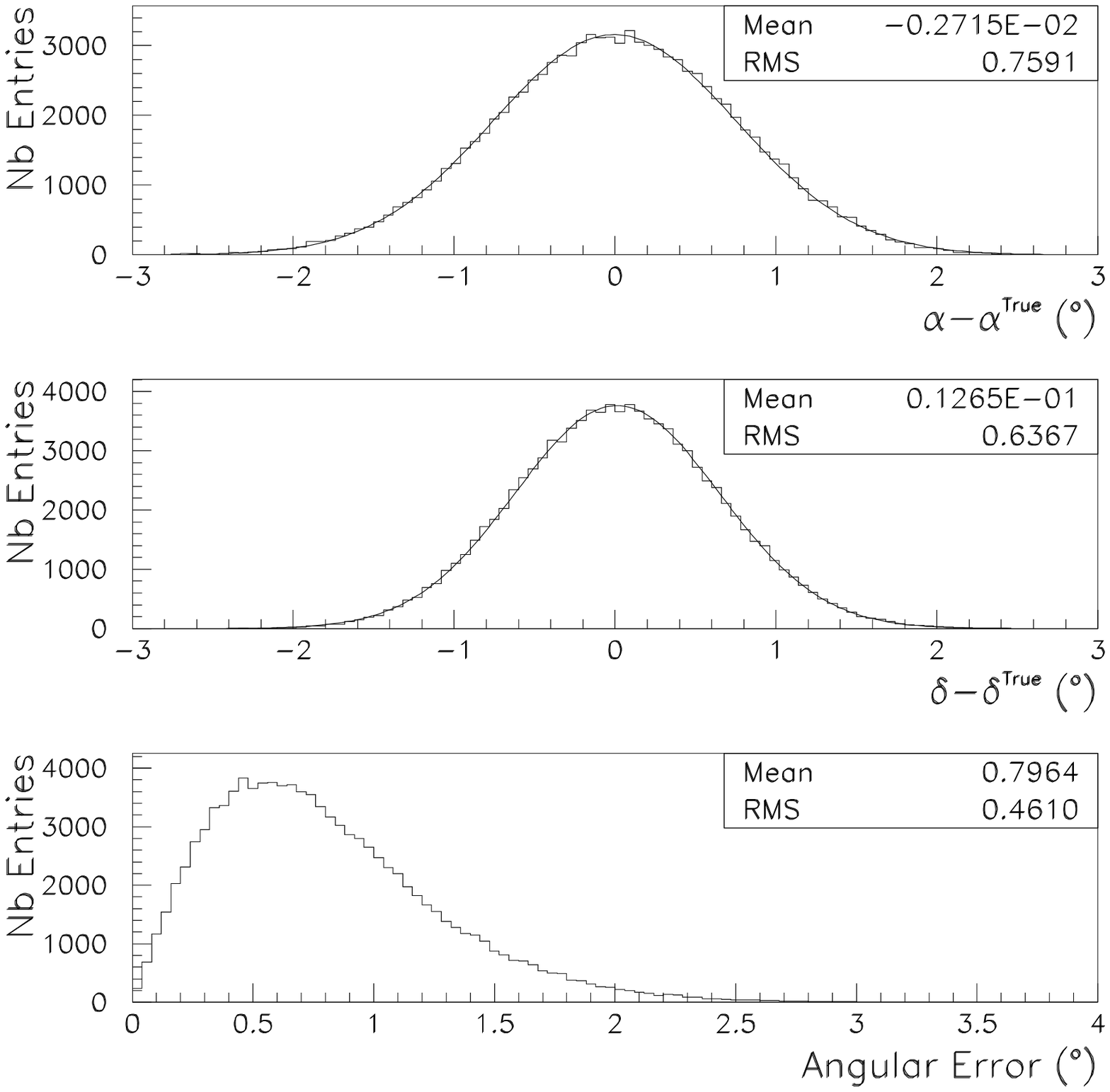}
  \caption{Reconstruction accuracy for a source at the Galactic Center at a fixed day time for LIGO-Virgo network.\\
The angular error is the angular distance on the sphere between the true direction and the reconstructed one.
The curves on the first two plots correspond to the best Gaussian fit.}
  \label{fig:reco_pos}
\end{figure}

In this section, we only deal with the LIGO-Virgo network and the effect of the antenna-pattern functions are not included and it is 
assumed that all detectors measure the arrival time with the same precision. 
As previously said, it allows to decouple the effect of the antenna-patterns and of the location with respect to the 
3-detector plane.

First of all, as an example, in order to evaluate the 
accuracy of the reconstruction, we choose a given position in the sky (coordinates of the Galactic Center
$\alpha_{GC}= 266.4^{\circ}, \delta_{GC}=-28.98^{\circ}$)
and we perform the simulation with $\sigma_i=10^{-4}$ s at a fixed time.
The results are shown on Figure \ref{fig:reco_pos}.
A resolution of about 0.7 degrees can be achieved both on $\alpha$ and $\delta$.
The angular error is defined as the angular distance on the sphere between the true direction and the reconstructed one
(it does not depend on the coordinate system and in particular there is no divergence (only due to the coordinate system)
when $\delta$ is equal to 90 degrees).
This variable will be used in the following steps as the estimator of the reconstruction quality.
The mean angular error is 0.8 degrees.
As shown on Table \ref{tab:covar}, the estimated errors (given by the covariance matrix) obtained by the $\chi^2$ minimization are in perfect 
agreement with these resolutions. 

\begin{table}[h]
  \centering
  \begin{tabular}{|c|c|c|}
\hline
Angle    & RMS of Distribution    & Mean of estimated errors given \\
         &                        & by the covariance matrix\\ \hline
$\alpha$ & 0.760$^{\circ}$        & 0.758$^{\circ}$\\ \hline
$\delta$ & 0.635$^{\circ}$        & 0.636$^{\circ}$\\ \hline
  \end{tabular}
  \caption{Reconstruction accuracies on $\alpha$ and $\delta$ and errors given by the covariance matrix. Three digits are given in order
to show the adequacy between RMS and errors given by the covariance matrix.}
  \label{tab:covar}
\end{table}

Varying $\sigma_i$ in the simulation, it is found that the resolution is proportional to the timing resolution
for reasonable errors ($\sigma_i \le 3$ ms. For $\sigma_i \gg 3$ ms, the errors become comparable to the sky size.) and we obtain:

\begin{eqnarray}
  \label{eq:reso_vs_sigma}
  \alpha_{Resolution} &=& 0.7^{\circ} ~~ \frac{\sigma_i}{10^{-4} \mbox{s}}\\ \nonumber
  \delta_{Resolution} &=& 0.6^{\circ} ~~ \frac{\sigma_i}{10^{-4} \mbox{s}} \\ \nonumber
  \mbox{Mean Angular error} &=& 0.8^{\circ} ~~ \frac{\sigma_i}{10^{-4} \mbox{s}}\nonumber
\end{eqnarray}

\begin{figure}
  \includegraphics[width=10cm]{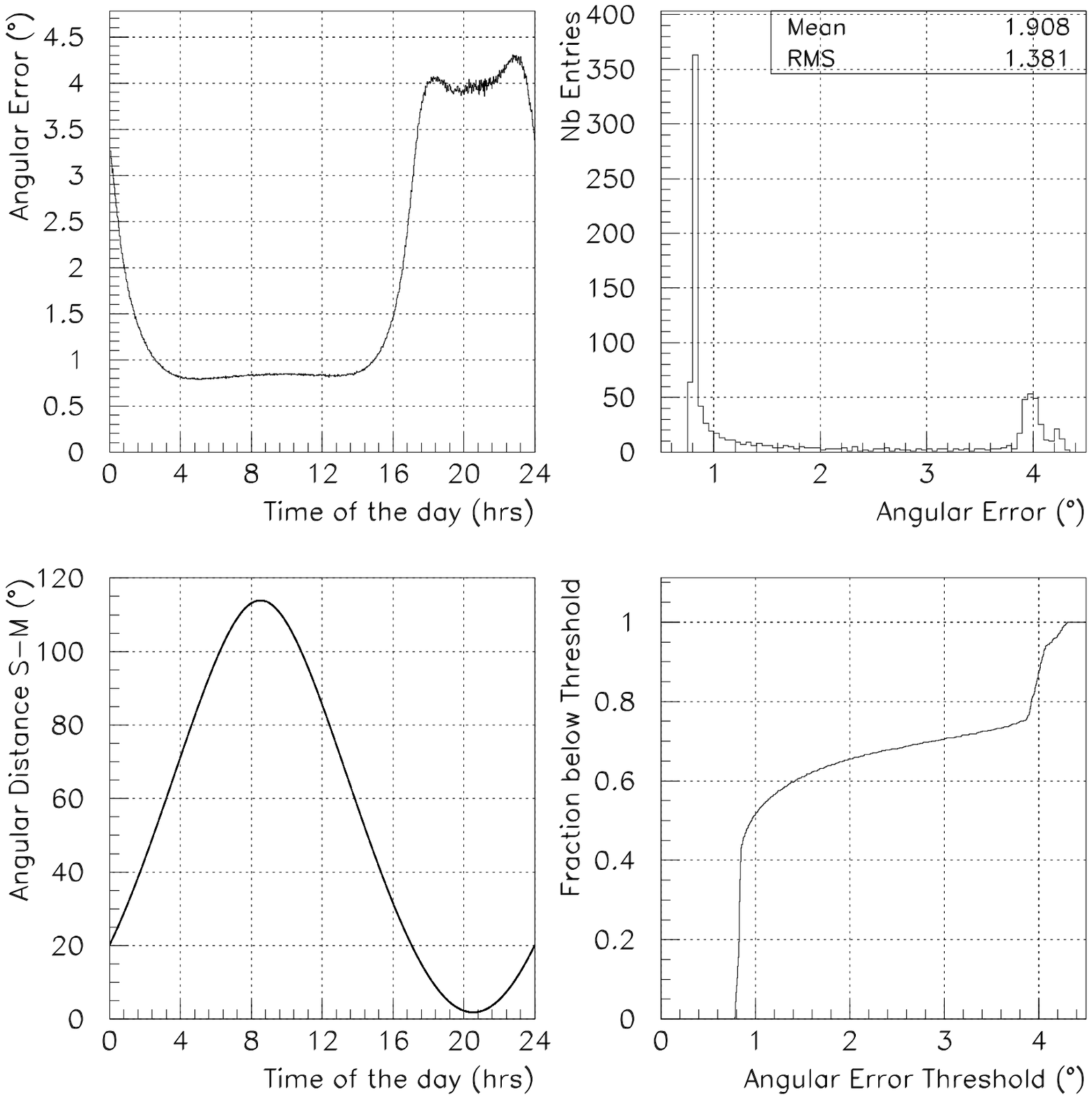}
  \caption{Angular Error for a source at the Galactic Center for one day (t=0 arbitrary chosen) for LIGO-Virgo network.\\ The top-left plot presents
the angular error as a function of time. 
The bottom-left plot shows the angular distance between the source (S) 
and its mirror image (M) (other direction in the sky which gives the same time delays).
The top-right plot is the distribution of angular errors.
The bottom-right gives the fraction of events with an angular error below a given threshold versus this threshold.
$\sigma_i$ is set to 0.1 ms for all detectors.}
  \label{fig:reco_day}
\end{figure}

For the Galactic Center direction and with $\sigma_i = 0.1$ ms, 
we evaluate the reconstruction performances over one entire day, still neglecting the antenna-pattern.
 
It clearly appears on Figure \ref{fig:reco_day} 
that the resolution varies, as expected, during the day and lies
between $0.8^{\circ}$ and $4.3^{\circ}$ (see Table \ref{tab:reso_nobp} for details).

\begin{table}
  \centering
  \begin{tabular}{|c|c|c|c|c|}
\hline
$\delta (^{\circ})$ & Minimal Error $(^{\circ})$& Maximal Error $(^{\circ})$& Mean Error $(^{\circ})$& Median Error $(^{\circ})$\\ \hline
 -28.98 (GC)      &  0.8 & 4.3 & 1.9 & 0.95 \\ \hline
  0               &  1.3 & 3.1 & 1.8 & 1.5 \\ \hline
All values        &  0.7 & 4.5 & 1.6 & 1.1 \\ \hline 
  \end{tabular}
  \caption{Angular resolution for Galactic Center, $\delta=0^{\circ}$ position and averaged on all possible values of
$\delta$ taken over the whole day. No antenna-pattern effect is included.}
  \label{tab:reso_nobp}
\end{table}

\begin{figure}
  \includegraphics[width=10cm]{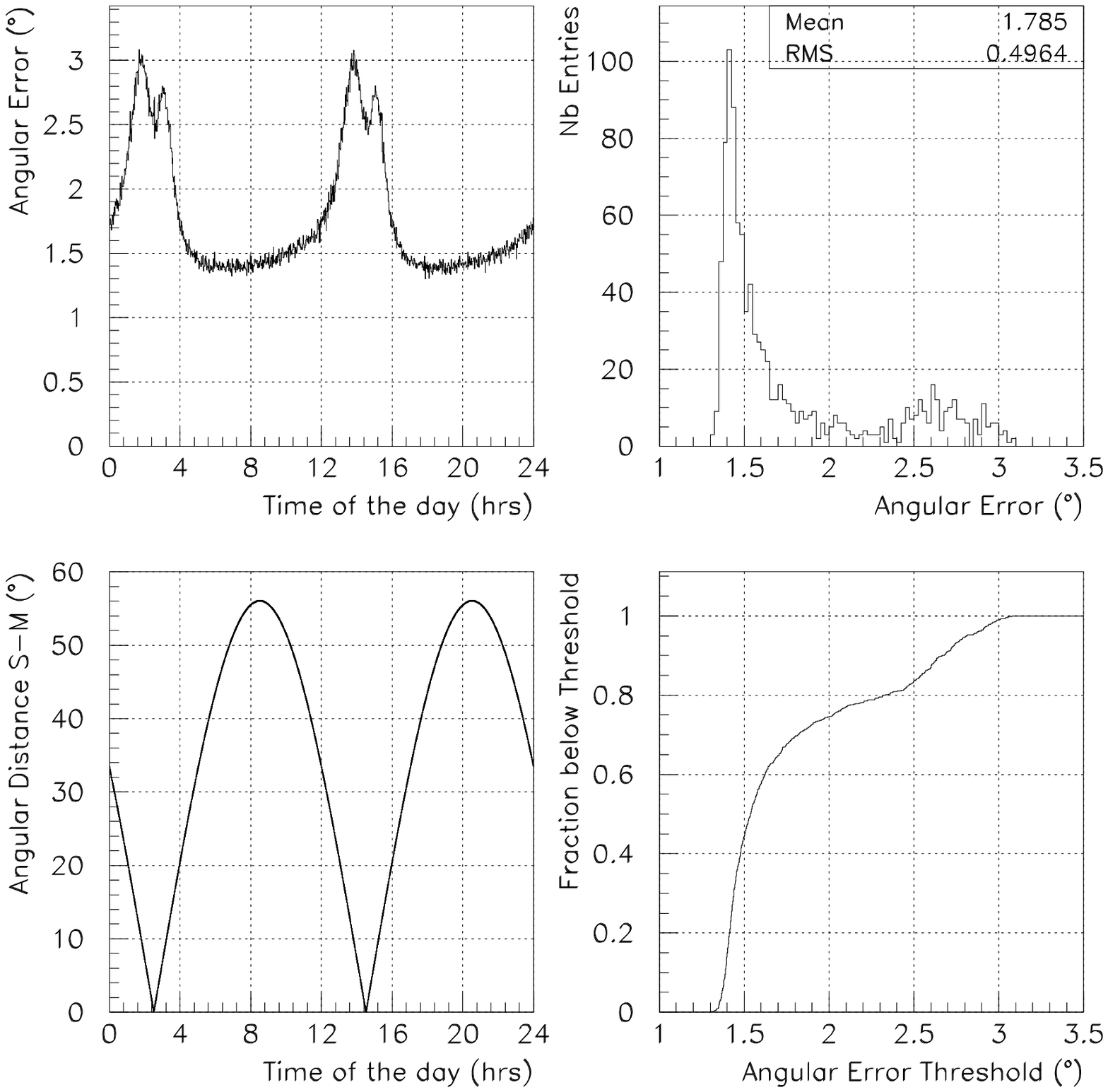}
  \caption{Angular error for a source at $\delta = 0^{\circ}$ for one day (t=0 arbitrary chosen) for LIGO-Virgo network.\\ The top-left plot presents
the angular error as a function of time. 
The bottom-left plot shows the angular distance between the source (S) 
and its mirror image (M) (other direction in the sky which gives the same time delays).
The top-right plot is the distribution of angular errors.
The bottom-right gives the fraction of events with an angular error below a given threshold versus this threshold.
$\sigma_i$ is set to 0.1 ms for all detectors.}
  \label{fig:reco_day2}
\end{figure}

The error maximum which appears around $t=20 $h (the time origin is arbitrary) corresponds to directions
for which the source approaches the detector plane. The angular distance between the two possible solutions 
(the source and its mirror image) gives a good estimator of this closeness (see
Fig. \ref{fig:reco_day}). This effect is even clearer on Figure \ref{fig:reco_day2}
where the source is located at $\delta = 0^{\circ}$. In this case, 
the source crosses the 3-detector plane twice a day ($t=2.4$ h, $t=14.4$ h)
and the angular distance between the real source and its mirror image is null.
This degradation of the angular reconstruction is not related to the $\chi^2$ minimization technique. It is
a geometrical property of the network which releases constraints when the source belongs to the 3-detector plane.
For the $\delta = 0^{\circ}$ configuration, angular resolutions as given in Table \ref{tab:reso_nobp} are similar
to the ones obtained for the Galactic Center.

This degradation can be easily understood with only two detectors located at $(\pm d/2,0,0)$ and a source in the
(x,y) plane defined by its angle $\theta$ with the x axis. In this 2-detector case, the arrival time difference $t_{21}$ is given by:
\begin{equation}
  \label{eq:2d_rel}
t_{21} = \frac{d}{c} \cos \theta 
\end{equation}

and thus, if $t_{21}$ is measured with an error $\delta t$, it will induce an error 
$\delta \theta = \frac{c \delta t}{d | \sin \theta |} $ on the source direction which diverges when the source belongs
to the 2-detector line ($\theta=0$ or $\pi$). Of course, this simple estimation of the angular error supposes that Eq.
\ref{eq:2d_rel}
between $t_{21}$ and $\theta$ can always be inverted, assumption which fails when $t_{21}$ becomes greater than $\frac{d}{c}$ due to
measurement errors. We can wonder how to handle this case. In the 2-detector example, we minimize $\chi^2$ defined by:

\begin{equation}
  \label{eq:chi2_2d}
  \chi^2(\theta) = \frac{1}{\delta t^2} \left( t_{21}^{measured} - \frac{d}{c} \cos \theta \right)^2
\end{equation}

which becomes minimal for $\theta = \mbox{acos}( t_{21}^{measured} \times \frac{c}{d} )$ if $|t_{21}^{measured} \times \frac{c}{d}|\le 1$ or
$\theta = 0$ when the previous condition is not satisfied. 
We observe similar effect in the 3-detector case when the source approaches the 3-detector plane.

\begin{figure}
  \includegraphics[width=10cm]{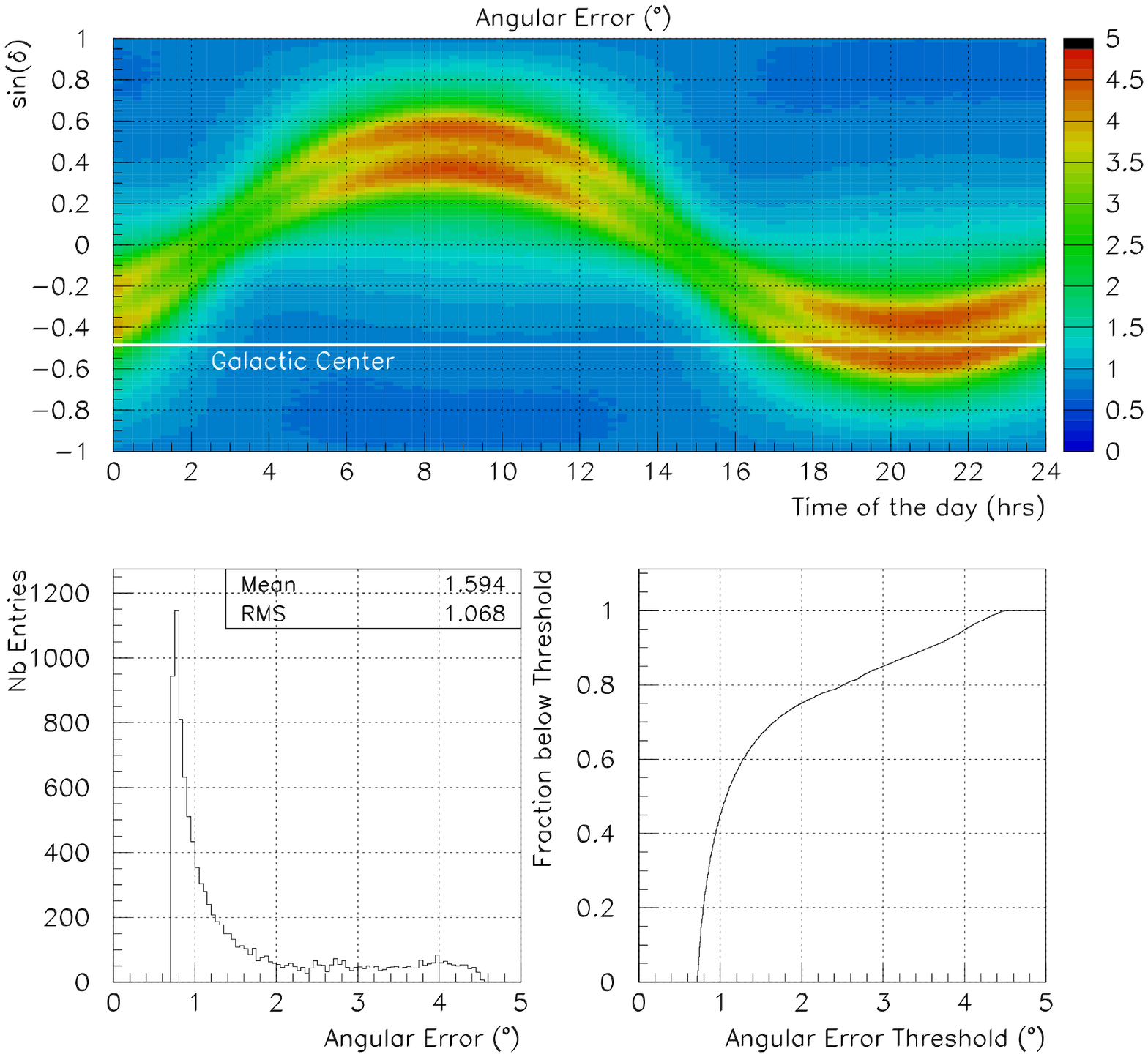}
  \caption{Reconstruction resolution for the whole sky for LIGO-Virgo network.\\ 
The top plot presents the angular error as a function of time for all possible values of $\delta$. The bottom-left plot
is the distribution of angular errors.
The bottom-right gives the fraction of events with an angular error below a given threshold versus this threshold.
The white line gives the trajectory of the Galactic Center
during the day.
$\sigma_i$ is set to 0.1 ms for all detectors.}
  \label{fig:reco_sky}
\end{figure}

Figure \ref{fig:reco_sky} presents the angular errors averaged on all possible value of $\delta$. In this case, 
the angular resolutions are given in Table \ref{tab:reso_nobp}, ranging
0.7 to 4.5 degrees.
Without considering antenna-pattern effect, 
it means that an angular error lower than about one degree can be reached for half of the sky
and 90 \% is below 3.5 degrees. The intersection of the 3-detector with the celestial sphere appears
as a zone with worst angular resolutions (error ranging from 3 degrees to 4.5 degrees)
of the source direction.

\section{Including the Antenna-Pattern effect in a 3-detector network}
\label{including}

As in previous section, we use the LIGO-Virgo network as benchmark.
Now, the strength of the signal seen by each detector is computed taking into account the
antenna-pattern functions. The errors on arrival times are estimated using Eq.\ref{eq:sigma_t}
with $\zeta=1$ and $\sigma_0= 1$ms.
In order to remain close to the case described in Section III (all timing errors equal to 0.1 ms),
we impose that the mean SNR over the three detectors is equal to 10. A linear polarization has been assumed for the incoming signal.

\begin{figure}
  \includegraphics[width=10cm]{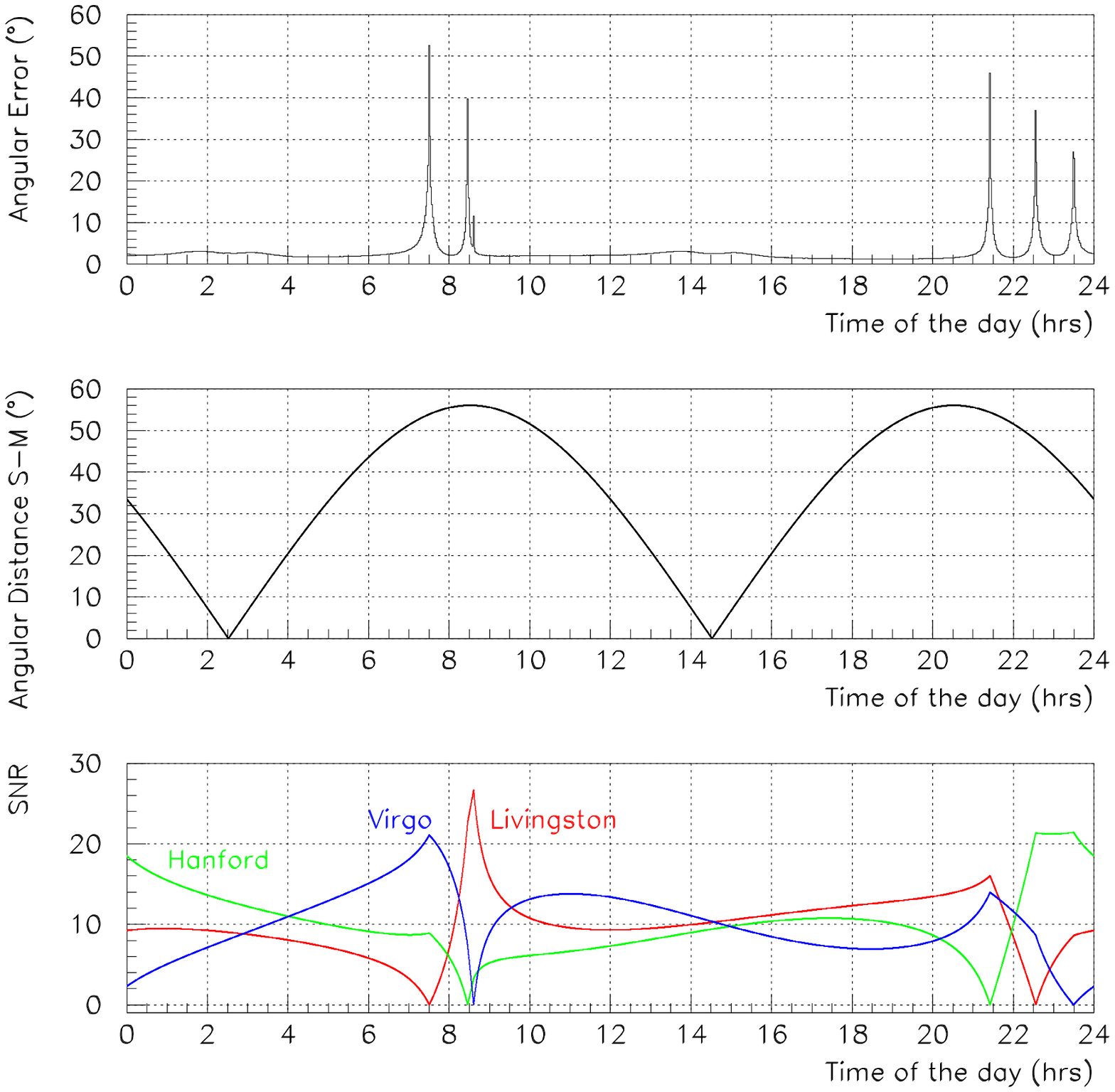}
  \caption{Angular Error for a source at $\delta = 0^{\circ}$ as a function of the day time (t=0 arbitrary chosen) for LIGO-Virgo network.\\ 
The top plot presents
the angular error as a function of time. The middle plot shows the angular distance between the source and its mirror 
image (the other sky direction giving the same time delays). The SNR values in each interferometer are plotted on the bottom
figure. The effect of the antenna-pattern functions is included and 
the mean SNR is set to 10.}
  \label{fig:reco_day3}
\end{figure}

Figure \ref{fig:reco_day3} presents the angular error as a function of the day time for a source located
at $\delta=0^{\circ}$ and can be compared to
Fig. \ref{fig:reco_day2}. The two broad peaks ($t=2.4$ hours and $t=14.4$ hours),
corresponding to the time when the source belongs to the 3-detector
plane, are still present (barely seen due to the change of scale) but six sharp diverging peaks show up. 
They correspond to blind regions for at least one detector, as expected.

\begin{figure}
  \includegraphics[width=10cm]{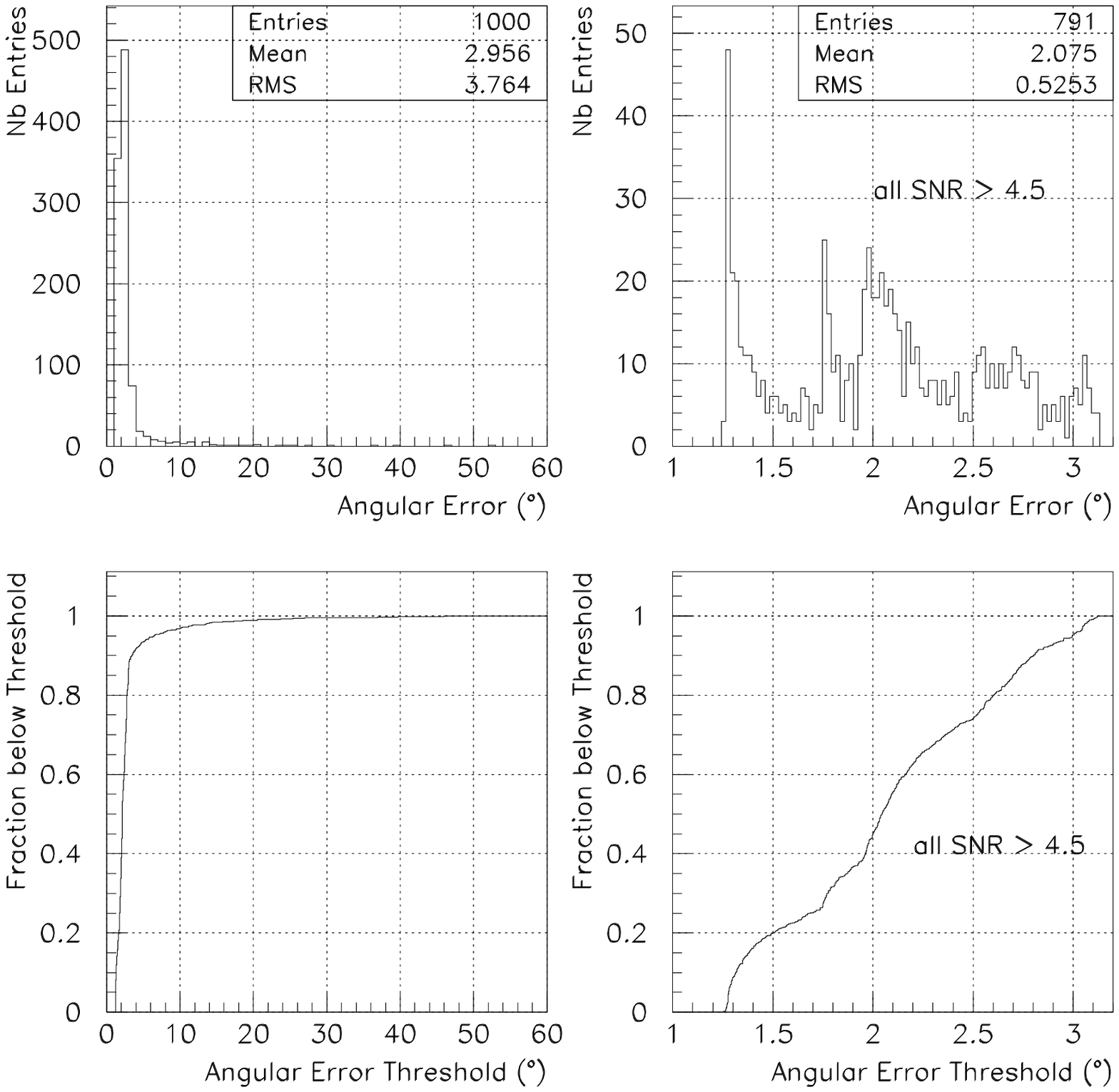}
  \caption{Distribution of angular errors for a source at $\delta = 0^{\circ}$ for LIGO-Virgo network.\\
The top-left plot shows the distribution of angular error for all events. 
The bottom-left one presents the fraction of events with an angular error below a given threshold versus this threshold.
The right plots are the same imposing that the SNR seen by each detector is greater than 4.5.}
  \label{fig:reco_day3bis}
\end{figure}

\begin{table}
  \centering
  \begin{tabular}{|c|c|c|c|c|}
\hline
$\delta (^{\circ})$ & Event Fraction(\%)& Min. Error $(^{\circ})$& Mean Error $(^{\circ})$& Median Error $(^{\circ})$\\ \hline
 -28.98 (GC)      & 100   & 0.7 & 4.0 &  1.8 \\ 
Any SNR & & & & \\\hline
 -28.98 (GC)      & 56.4      &  0.7 & 1.8 &  1.25 \\
All SNR $\ge 4.5$ & & & & \\\hline 
  0               & 100      &1.2 & 3. &  2.2 \\ 
 Any SNR& & & & \\\hline
  0               &  79.1     &1.2 & 2.1 & 2. \\ 
All SNR $\ge$ 4.5& & & & \\\hline
All values        &  100      &0.7 & 2.7 &  1.7 \\ 
Any SNR& & & & \\\hline
All values        &  59.8         &  0.7 & 1.8 & 1.3 \\ 
All SNR $\ge$ 4.5& & & & \\\hline
\end{tabular}
  \caption{Angular resolution for Galactic Center, $\delta=0^{\circ}$ position and averaged on all possible values of
$\delta$ taken over the whole day. The antenna-pattern effect is included. The column ``Event Fraction'' gives the fraction of events
satisfying the SNR selection criterion.}
  \label{tab:reso_bp}
\end{table}

Of course, in real conditions, a minimal threshold will be applied for the event selection. For this purpose,
 we define a SNR threshold equal to 4.5 on each 
detector which leads (supposing a Gaussian noise) to a false alarm rate about $10^{-6}$ Hz when requiring a triple coincidence between Virgo, 
LIGO-Hanford and 
LIGO-Livingston. About 79 \% of the events satisfy this constraint and a median error of $2.0^{\circ}$ is reached 
(See Table \ref{tab:reso_bp} and Fig. \ref{fig:reco_day3bis}).

\begin{figure}
  \includegraphics[width=10cm]{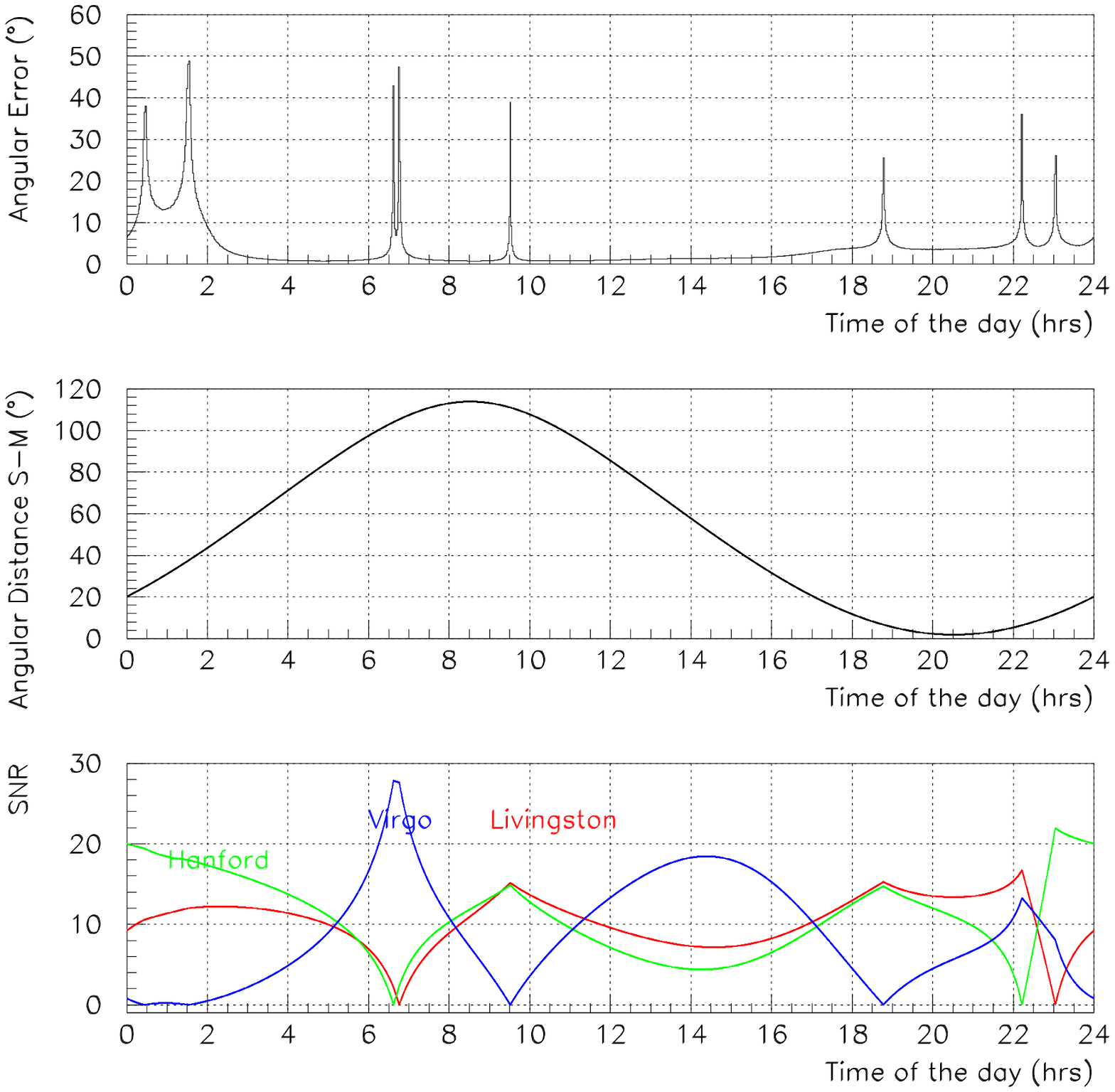}
  \caption{Angular Error for a source at the Galactic Center as a function of the day time (t=0 arbitrary chosen) for LIGO-Virgo network.\\ 
The top plot presents
the angular error as a function of time. The middle plot shows the angular distance between the source and its mirror 
image (the other sky direction giving the same time delays). The values of the antenna-pattern functions are plotted 
on the bottom figure. The effect of the antenna-pattern functions is included and 
the mean SNR is set to 10.}
  \label{fig:reco_day4}
\end{figure}

For a source located at the Galactic Center, the angular error shows several spikes related to the antenna-pattern effect and 
the geometry of the network only
increases the mean error in the region around $t=21$ h (Fig. \ref{fig:reco_day4}). All peaks correspond
to regions which are blind for at least one interferometer. Imposing that all SNR are above 4.5 leaves 56.4 \% of the events
and in particular all these blind regions are removed.
The obtained angular resolutions are given in Table 
\ref{tab:reso_bp}. They are slightly better than for the $\delta=0^{\circ}$ case.

\begin{figure}
  \includegraphics[width=10cm]{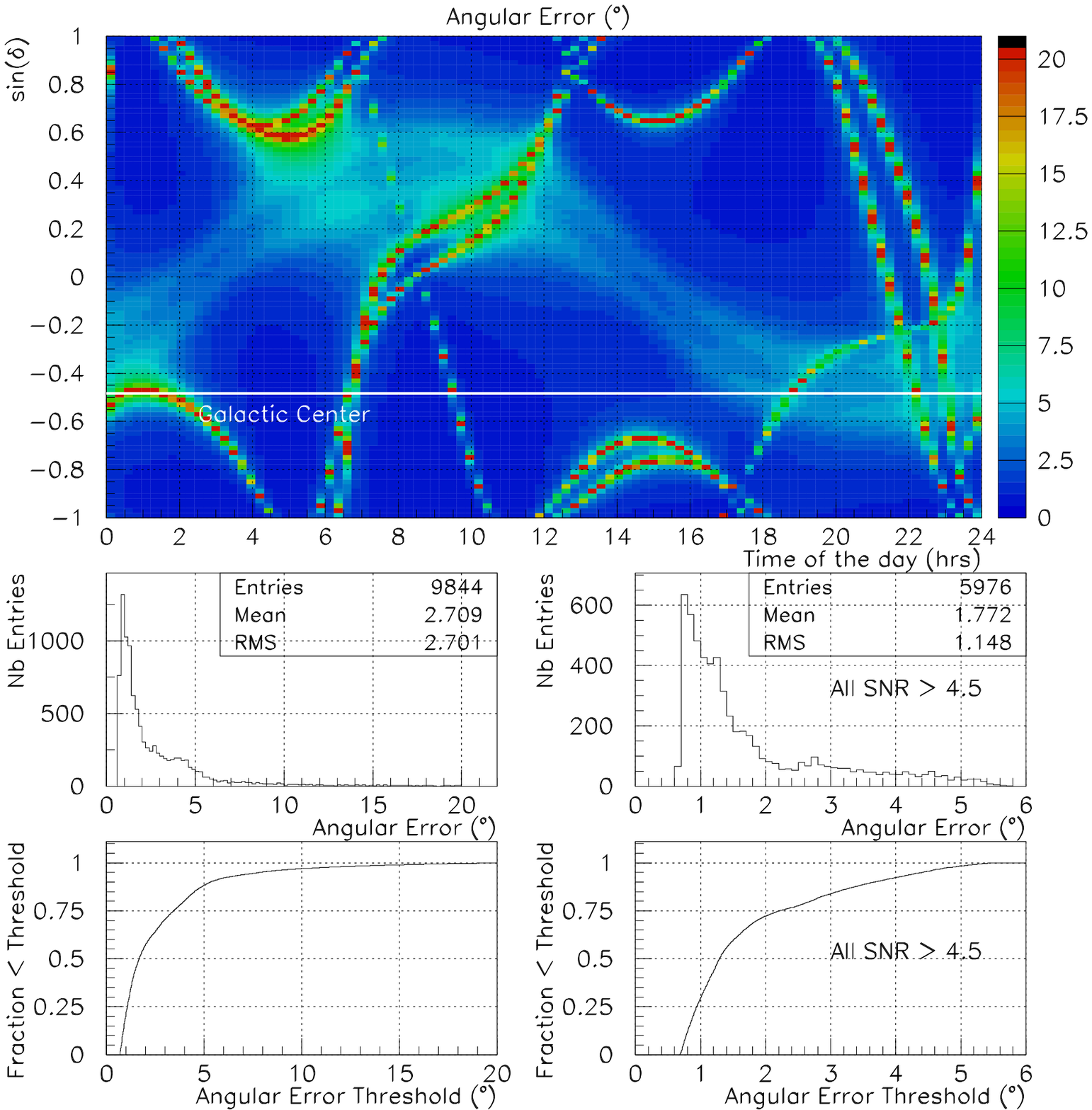}
  \caption{Reconstruction resolution for the full sky for LIGO-Virgo network. 
The top graph gives the angular error as a function of $\alpha$ and $\sin(\delta)$. For clarity, the 
values above $20^{\circ}$ have been set to $20^{\circ}$. The bottom plots presents the angular error
distribution and the fraction of events below a given angular threshold imposing or not the SNR condition
(all SNR$>$4.5).}
  \label{fig:reco_skybp}
\end{figure}

Figure \ref{fig:reco_skybp} presents the angular errors averaged on all possible
$\delta$. As in Fig. \ref{fig:reco_sky}, the intersection between the 3-detector plane  
and the celestial sphere is visible. The regions of largest errors ($> 15^{\circ}$)
corresponds to the blind regions of the various detectors. Without any conditions on the SNR in each detector,
we obtain a median error of $1.7^{\circ}$ (see Table \ref{tab:reso_bp}). 
The SNR condition is satisfied for 60\% of the events and
leads to a median error of $1.3^{\circ}$ with 90 \% of the events below $3.7^{\circ}$.
In the most favorable cases, we can even reach a precision of $0.7^{\circ}$. A resolution better than $1^{\circ}$ is obtained
in 30\% of the cases.

\section{Addition of other gravitational wave detectors}

First of all, the 2-kilometer long Hanford interferometer has been included in the network. In this case, we supposed that the
SNR seen by by Hanford 2k is half of the SNR seen by Hanford 4k. As expected, adding Hanford 2k does not change
significantly the resolutions obtained in previous sections.

\begin{figure}
  \includegraphics[width=10cm]{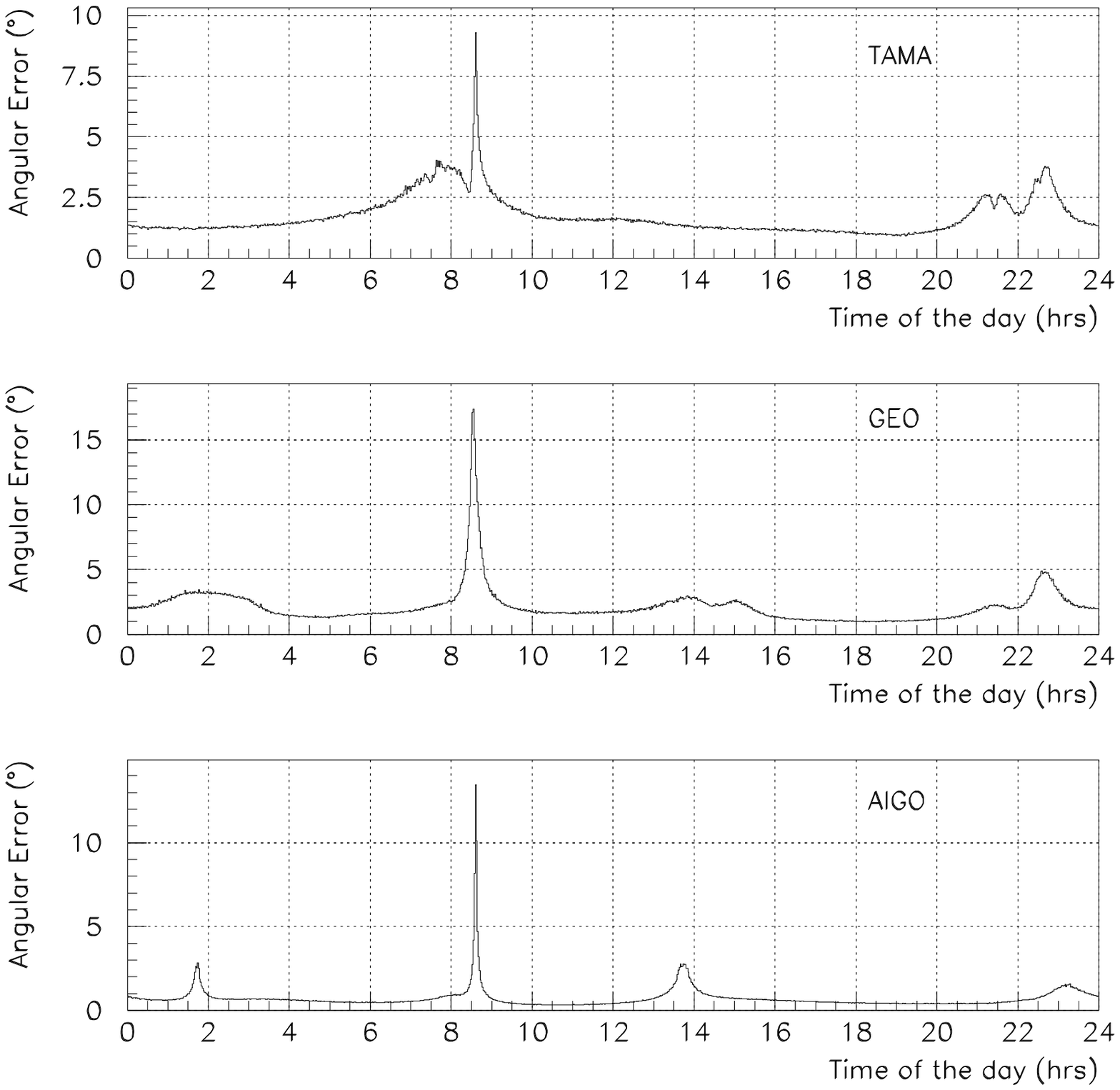}
  \caption{Reconstruction resolution for a source at $\delta = 0^{\circ}$ as a function of the day time (t=0 arbitrary chosen) 
adding a fourth detector to the LIGO-Virgo network.\\
The top figure presents the angular error as a function of time adding TAMA, the middle one adding GEO and the bottom one adding AIGO.}
  \label{fig:reco_00bp_4itf}
\end{figure}

Then, we add TAMA \cite{TAMA}, GEO \cite{GEO} and AIGO \cite{AIGO} assuming (for sake of simplicity) the same sensitivities as LIGO and Virgo. We still
impose a mean SNR equal to 10.
Figure \ref{fig:reco_00bp_4itf}, which has to be compared to Figure \ref{fig:reco_day3}, shows the angular reconstruction for a source
located at $\delta = 0^{\circ}$ as a function of the day time adding only a fourth detector to the LIGO-Virgo network.
Of course, there is no longer an ambiguity in the possible solutions and
all former blind regions are attenuated except the one around $t=8.5 h$ for which effective low SNRs are obtained with both Hanford and Virgo.
The mean error ($3^{\circ}$ for LIGO-Virgo Table \ref{tab:reso_bp}) is lowered to $2.0^{\circ}$ adding GEO, to $1.7^{\circ}$ adding TAMA
and $0.7^{\circ}$ adding AIGO. Not surprisingly, the larger the network baseline, the better the resolution.

\begin{figure}
  \includegraphics[width=10cm,height=7cm]{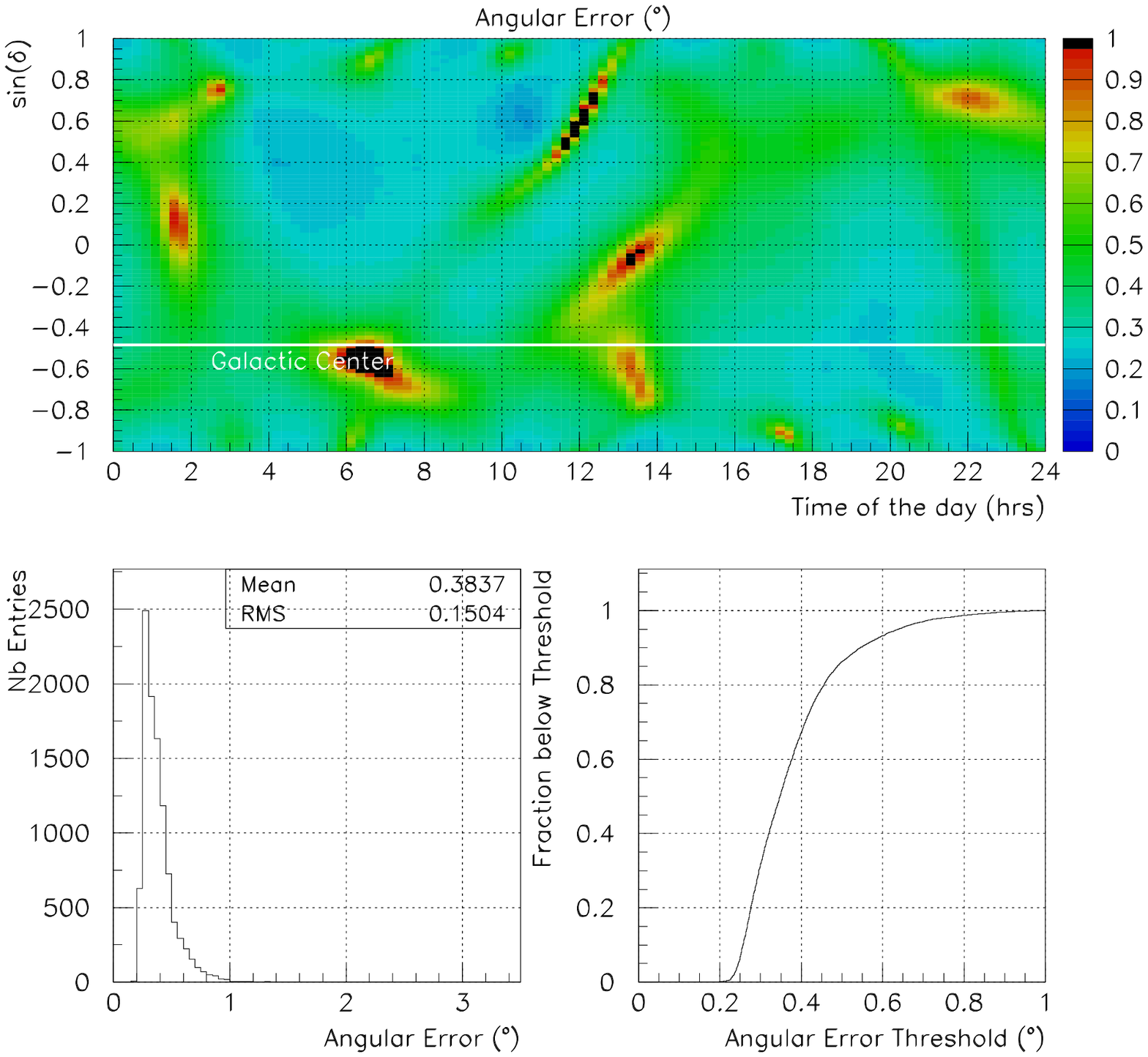}
  \caption{Reconstruction resolution for the full sky with a network of 6 detectors.
The top graph gives the angular error as a function of $\alpha$ and $\sin(\delta)$.For clarity, the 
values above $1^{\circ}$ have been set to $1^{\circ}$.The bottom plots presents the angular error
distribution and the fraction of events below a given angular threshold.}
  \label{fig:reco_sky_6itf}
\end{figure}

In order to evaluate the performances of a strongly overconstrained network, we evaluate the reconstruction resolution for the full sky
using the 6 detectors (still using the same sensitivity). The results are shown in Figure \ref{fig:reco_sky_6itf} similar to 
Figure \ref{fig:reco_skybp}.
The mean error is about $0.4^{\circ}$ and  99\% of the sky is covered with a resolution below $0.8^{\circ}$. The regions with errors between 
1 and 3 degrees correspond to directions for which several interferometers have very low SNRs (as the region around $t=8.5 h$ in the previous 
figure).

\section{Effects of Systematic Errors on Arrival Times}

\begin{table}
  \centering
  \begin{tabular}{|c|c|c|c|}
\hline
Bias (ms)   & Bias on $\alpha (^{\circ})$    & Bias on $\delta(^{\circ})$ & Angular Bias $(^{\circ})$\\\hline
.1          &  -0.24                         &  0.65                      & 0.68 \\ \hline
.2          &  -0.48                         &  1.26                      & 1.33 \\ \hline
.3          &  -0.72                         &  1.79                      & 1.90 \\ \hline
.4          &  -0.97                         &  2.55                      & 2.69 \\ \hline
.5          &  -1.2                          &  3.17                      & 3.35 \\ \hline
1.          &  -2.6                          &  6.20                      & 6.63 \\ \hline
  \end{tabular}
  \caption{Effect of a timing bias on angular reconstruction.\\
The bias is applied to the arrival time at Livingstone interferometer. 
The RMS of the statistical errors has been set to .1 ms. For this direction, the statistical angular error is about 0.8$^{\circ}$.}
  \label{tab:bias}
\end{table}

All angular errors quoted previously suppose that arrival time measurements are only subject
to Gaussian noise. Systematic biases can also be introduced by the analysis and their effect can
be evaluated. In order to do so, we modify Equation \ref{eq:t_measured_bias}
introducing a timing bias for only one detector:

\begin{equation}
  \label{eq:t_measured_bias}
  t_i^{Measured} = t_i^{true} + GaussianRandom \times \sigma_i + Bias
\end{equation}

As in sections \ref{source} and \ref{including}, we only consider the LIGO-Virgo network.
It appears that the widths of the distribution for reconstructed $\alpha$ and $\delta$ are not modified by the bias but the central values are
shifted from the true ones. The differences between the reconstructed value and the true one are proportional to the bias and are 
significantly different from zero when the bias and the statistical error have the same order of magnitude. Table \ref{tab:bias} shows
the effect of the bias for a given direction (we check that the effect is independant of the source location). In this example, the bias has been
applied to the Livingstone interferometer. The width of statistical errors on arrival time was .1 ms leading to a statistical angular error about 
 0.8$^{\circ}$.
For the tested configurations, we do not observe significant differences between the three interferometers of the network.

\section{Conclusion}

We described a method for the reconstruction of the source direction using the timing information (arrival time and 
associated error)
delivered by gravitational wave detectors such as LIGO and Virgo. The reconstruction is performed
using a least-square minimization which allows to retrieve the angular position of the source and
the arrival time at the center of the Earth. The minimization also gives an estimation of errors
and correlations on fitted variables. For a given position, the angular error is proportional to the timing 
resolution and the systematic errors (if they exist) introduce a significant bias on reconstructed angles when they reach the
level of the statistical one.

When the antenna-pattern effect is included and imposing a mean SNR
value of 10 in the LIGO-Virgo network, a precision of $1.7^{\circ}$ can be reached for half of the sky.
In order to reproduce a realistic case, we apply a threshold on the SNR in each detector (SNR$>$ 4.5 leading
to a false alarm rate about $10^{-6}$ Hz when performing a threefold coincidence). This condition is 
satisfied for 60 \% of the sky and the median angular error in this case is $1.3^{\circ}$.
As a resolution of $1^{\circ}$ is obtained for 30 \% of the events satisfying the SNR condition, it means 
that about 20 \% of the whole sky is seen with an angular error lower than $1^{\circ}$.

Adding other gravitational waves detectors allows to reduce the blind regions and to lower the mean resolution. In the best considered case
(6 detectors), the resolution is about $0.4^{\circ}$ and 99\% of the sky is seen with a resolution lower than $0.8^{\circ}$.

All quoted resolutions (about one degree) are similar to those delivered by $\gamma$-ray satellites when the first
GRB counterparts have been identified. So, we can expect it will be also sufficient for the first identification of gravitational wave
sources.

\end{document}